\documentclass[12pt,a4paper]{article}
\pdfoutput=1
\usepackage[utf8]{inputenc}
\usepackage{textcomp}
\usepackage{macros}

\preprint{}
\title{Effective Bethe Ansatz for Spin-1 Non-integrable Models}

\author[a] {Zhuohang Wang,}
\author[a,b]{Rui-Dong Zhu}
\affiliation[a]{Institute for Advanced Study \& School of Physical Science and Technology,\\ Soochow University, Suzhou 215006, China}
\affiliation[b]{Jiangsu Key Laboratory of Frontier Material Physics and Devices,\\ Soochow University, Suzhou 215006, China}

\abstract{This work presents a comprehensive benchmark and validation of a recently proposed method called Effective Bethe Ansatz (EBA). It is a variational method that deforms the exact Bethe wavefunctions of one-dimensional spin chains at integrable points to approximate non-integrable systems. We apply this method to the non-integrable regime of the spin-1 bilinear-biquadratic chain. By performing EBA method starting from the two integrable endpoints, the Takhtajan–Babujian point and the Lai–Sutherland point, we systematically evaluate the accuracy of the EBA for the ground state and first excited state. Our validation is based on a direct comparison with exact diagonalization, assessing energy, fidelity, and entanglement entropy. The results confirm that the EBA provides a quantitatively accurate description in a finite window around the integrable points, while its fidelity and entanglement properties degrade in a controlled way as the perturbation increases. The method successfully captures key finite-size effects, such as level crossings, manifested as sharp drops in fidelity, and provide a probe to potential phase transitions. This study establishes the EBA as a reliable and efficient semi-analytical tool, clarifying its scope and limitations for studying low-energy physics in non-integrable quantum spin chains.}

\begin{document}

\allowdisplaybreaks

\maketitle

\section{Introduction}

Quantum spin chains serve as a paradigm for studying strongly correlated quantum many-body physics, offering profound insights into quantum magnetism, topological phases, and quantum criticality. Integrable models, such as the XXX Heisenberg chain, are of particular interest as they admit exact solutions via the Bethe ansatz \cite{Bethe:1931hc}, enabling complete analytical understanding. In contrast, realistic systems are usually non-integrable, placing them in a regime where exact solutions are unavailable. Investigating these regimes necessitates reliance on relatively expensive numerical techniques like exact diagonalization (ED) or the density matrix renormalization group (DMRG), or on approximate analytical methods.

Recently, an approximate semi-analytical framework known as the Effective Bethe Ansatz (EBA) has been proposed to extend the Bethe ansatz wavefunction formalism to non-integrable models \cite{wenlong}. The core idea is to retain the functional form of the Bethe ansatz wavefunction even away from integrability, but allow the corresponding Bethe roots to shift from their original values due to perturbations. These ``effective'' Bethe roots are determined by optimizing a suitable loss function, such as the energy expectation value. This method has shown promise in spin-1/2 systems, efficiently providing a physically intuitive and relatively accurate approximation for the spectra and wavefunctions of non-integrable models.

In this work, we study the following one-dimensional spin-1 Hamiltonian,
\begin{equation}
    H=\frac{1}{4}\left[\sum_{i=1}^L\vec{S}_i\cdot \vec{S}_{i+1}+\beta \left(\vec{S}_i\cdot \vec{S}_{i+1}\right)^2\right],\label{ham}
\end{equation}
where the overall factor $1/4$ is simply a conventional choice adopted in the integrability context. This family of models harbors two well-known integrable points. At $\beta=-1$, the system corresponds to the integrable Takhtajan–Babujian model \cite{Babujian:1982ib,Takhtajan:1982jeo}, solvable by a standard (rank-1) Bethe ansatz. At $\beta=1$, it corresponds to the integrable Lai–Sutherland model \cite{Lai1974,sutherland-su3}, whose exact solution requires a nested Bethe ansatz first proposed in \cite{Yang:1967bm}. Notably, at $\beta=\frac{1}{3}$, the model maps exactly to the Affleck-Kennedy-Lieb-Tasaki (AKLT) model \cite{Affleck:1987vf,Affleck:1987cy}, which possesses an exactly solvable valence bond solid ground state with a gap, and is a cornerstone for understanding symmetry-protected topological phases. The model with general $\beta$ is also referred to as the Bilinear-Biquadratic (BLBQ) model in the literature. 

 In this article, we apply the Effective Bethe Ansatz method to the spin-1 model \eqref{ham} in the non-integrable region. For the first time, we apply the EBA approach to a spin-1 system requiring a nested Bethe ansatz framework and implement a bidirectional validation by extending the method from two distinct integrable endpoints. In particular, we extend from both endpoints, $\beta=-1$ (using the rank-1 ansatz) and $\beta=1$ (using the nested ansatz), into the non-integrable bulk ($-1<\beta<1$), focusing on the antiferromagnetic ground state and the first excited state. We perform a comprehensive evaluation of its accuracy and validity by comparing the EBA results with exact diagonalization for energy, fidelity, and entanglement entropy. Our calculations demonstrate that the EBA provides a good approximation to the true eigenstates near the integrable points, with accuracy degrading as one moves further away, consistent with theoretical expectations. 

The paper is organized as follows. In Section \ref{s:ABA-review}, we briefly review the Bethe ansatz solutions for the spin-1 model at the two integrable points ($\beta=\pm 1$). Section \ref{s:algorithm} details the Effective Bethe Ansatz methodology and the optimization algorithm for non-integrable models. Our main numerical results, obtained from extending from both endpoints, are presented and discussed in Section \ref{s:result}. Finally, we conclude and provide an outlook in Section \ref{s:conclusion}.

\section{Bethe Ansatz for Spin-1 Integrable Models}\label{s:ABA-review}

There are two integrable points in the family of spin-1 models \eqref{ham}: $\beta=\pm1$, but the Bethe ansatz takes rather different forms in these cases. 

At $\beta=-1$, one can use a rank-1 Bethe ansatz and take the spin-1 representation of the Lax matrix, 
\begin{equation}
    L_{0n}(u)=\left(\begin{array}{cc}
    \left( u+\frac{i}{2}\right)\mathbb{I}+iJ_n^z & iJ_n^-\\
    iJ_n^+ & \left( u+\frac{i}{2}\right)\mathbb{I}-iJ_n^z\\
    \end{array}
    \right),
\end{equation}
where $\vec{J}_n$ are spin-1 Pauli matrices at $n$-th lattice site and are given by
\begin{align}
    J^x=\left(\begin{matrix}
        0 & \sqrt{2} & 0\\
        \sqrt{2} & 0 & \sqrt{2}\\
        0 & \sqrt{2} & 0\\
    \end{matrix}\right),\quad J^y=\left(\begin{matrix}
        0 & -\sqrt{2}i & 0\\
        \sqrt{2}i & 0 & -\sqrt{2}i\\
        0 & \sqrt{2}i & 0\\
    \end{matrix}\right),\quad J^z=\left(\begin{matrix}
        2 & 0 & 0\\
        0 & 0 & 0\\
        0 & 0 & -2\\
    \end{matrix}\right),
\end{align}
and the spin operators are given by $\vec{S}:=\frac{1}{2}\vec{J}$. 
By defining the monodromy matrix, 
\begin{equation}
    T(u):=\prod_{i=1}^LL_{0i}(u)=\left(\begin{matrix}
        A(u) & B(u)\\
        C(u) & D(u)
    \end{matrix}\right),
\end{equation}
we decompose it on the $0$-th auxiliary space to obtain operator-valued entries $A(u)$, $B(u)$, $C(u)$ and $D(u)$. In particular, $B(u)$ and $C(u)$ are respectively the raising and lowering operators, and they respectively create or annihilate a quasi-particle called magnon in the spin system. One can solve the system with the Bethe ansatz, 
\begin{equation}
    \ket{\psi_M(\vec{u})}=\prod_{i=1}^MB(u_i)\ket{\Omega},\quad \ket{\Omega}:=\bigotimes_{i=1}^L\ket{\uparrow}_i=\left(\begin{matrix}
        1\\
        0\\
        0\\
    \end{matrix}\right)^{\otimes L}.\label{rank-1-ansatz}
\end{equation}
$u_i$'s are called the Bethe roots, and they satisfy the Bethe ansatz equation, 
\begin{equation}
    \frac{(u_j+i/2+i)^L}{(u_j+i/2-i)^L}=-\prod_{k=1}^M\frac{u_j-u_k+i}{u_j-u_k-i}.\label{BAE}
\end{equation}
We denote the normalized Bethe state as $\ket{\overline{\psi_M(\vec{u})}}$. 
In this article, we mainly focus on the anti-ferromagnetic case, with a unique ground state described by $L=M$. 

The other integrable point is located at $\beta=1$. Its Bethe ansatz is given in a rather different way, the nested Bethe ansatz. We need to use a $9\times 9$ R-matrix, 
\begin{equation}
    R(u)=u\mathbb{I}_{9\times 9}+{\cal P},
\end{equation}
to build the conserved charges of the system, where ${\cal P}$ denotes the permutation operator, 
\begin{equation}
    {\cal P}=\sum_{i,j=1}^3E_{ij}\otimes E_{ji},\quad (E_{ij})_{mn}=\delta_{i,m}\delta_{j,n}.
\end{equation}
To construct the wavefunction ansatz, we shall decompose the monodromy matrix $\tau(u):=\prod_{i=1}^LR(u)$ into 
\begin{equation}
    \tau(u)=\left(\begin{matrix}
        {\cal A}(u) & {\cal B}^{(1)}_{1}(u) & {\cal B}^{(2)}_{2}(u)\\
        {\cal C}^{(1)}_{1}(u) & {\cal D}_{11}(u) & {\cal B}_2(u)\\
        {\cal C}^{(1)}_{2}(u) & {\cal C}_2(u) & {\cal D}_{22}(u)\\
    \end{matrix}\right),
\end{equation}
then the nested Bethe state is given by 
\begin{equation}
    \ket{\Psi_{M_1,M_2}(\vec{u},\vec{v})}=\sum_{i_1,i_2,\dots,i_{M_1}=1}^2F_{i_1i_2\dots i_{M_1}}(\vec{v}){\cal B}^{(1)}_{i_1}(u_1){\cal B}^{(1)}_{i_2}(u_2)\dots {\cal B}^{(1)}_{i_{M_1}}(u_{M_1})\ket{\Omega},\label{nested-ansatz}
\end{equation}
where 
\begin{equation}
    F_{i_1i_2\dots i_{M_1}}(\vec{v}):=\bra{i_1i_2\dots i_{M_1}}{\cal B}_2(v_1){\cal B}_2(v_2)\dots {\cal B}_2(v_{M_2})\ket{1}^{\otimes M_1},
\end{equation}
\begin{equation}
    \ket{1}=\left(\begin{matrix}
        0\\
        1\\
        0\\
    \end{matrix}\right),\quad \ket{2}=\left(\begin{matrix}
        0\\
        0\\
        1\\
    \end{matrix}\right),\quad \bra{i_1i_2\dots i_{M_1}}=\bigotimes_{j=1}^{M_1}\bra{i_j}.
\end{equation}
Here ${\cal B}^{(1)}$ and ${\cal B}_2$ create quasiparticles associated with the two nested levels of the SU(3)-type Bethe ansatz. We denote the normalized nested Bethe state as $\ket{\overline{\Psi_{M_1,M_2}(\vec{u},\vec{v})}}$. 
The Bethe roots $\vec{u}$ and $\vec{v}$ are determined by the nested Bethe ansatz equations. 

Thus, we see that the Hamiltonian \eqref{ham} admits two distinct integrable descriptions, one based on a rank-1 Bethe ansatz at $\beta=-1$ and one on a nested Bethe ansatz at $\beta=1$.

\section{Effective Bethe Ansatz for Spin-1 Models beyond Integrability}\label{s:algorithm}

\begin{figure}
    \centering       
\begin{tikzpicture}[x=0.75pt,y=0.75pt,yscale=-1,xscale=1]
\draw    (75,119) -- (617,119) ;
\draw [shift={(620,119)}, rotate = 180] [fill={rgb, 255:red, 0; green, 0; blue, 0 }  ][line width=0.08]  [draw opacity=0] (8.93,-4.29) -- (0,0) -- (8.93,4.29) -- cycle    ;
\draw    (90,107) -- (90,131) ;
\draw    (570,108) -- (570,132) ;
\draw    (330,108) -- (330,132) ;
\draw    (410,108) -- (410,132) ;
\draw    (91,98) -- (199,98) ;
\draw [shift={(201,98)}, rotate = 180] [color={rgb, 255:red, 0; green, 0; blue, 0 }  ][line width=0.75]    (10.93,-3.29) .. controls (6.95,-1.4) and (3.31,-0.3) .. (0,0) .. controls (3.31,0.3) and (6.95,1.4) .. (10.93,3.29)   ;
\draw    (514,100) -- (567,100) ;
\draw [shift={(512,100)}, rotate = 0] [color={rgb, 255:red, 0; green, 0; blue, 0 }  ][line width=0.75]    (10.93,-3.29) .. controls (6.95,-1.4) and (3.31,-0.3) .. (0,0) .. controls (3.31,0.3) and (6.95,1.4) .. (10.93,3.29)   ;
\draw (611,132) node [anchor=north west][inner sep=0.75pt]   [align=left] {$\displaystyle \beta $};
\draw (79,144) node [anchor=north west][inner sep=0.75pt]   [align=left] {$\displaystyle -1$};
\draw (564,142) node [anchor=north west][inner sep=0.75pt]   [align=left] {$\displaystyle 1$};
\draw (324,142) node [anchor=north west][inner sep=0.75pt]   [align=left] {$\displaystyle 0$};
\draw (391,143) node [anchor=north west][inner sep=0.75pt]   [align=left] {AKLT};
\draw (88,73) node [anchor=north west][inner sep=0.75pt]   [align=left] {rank-1 effective Bethe ansatz};
\draw (431,75) node [anchor=north west][inner sep=0.75pt]   [align=left] {nested effective Bethe ansatz};
\end{tikzpicture}
    \caption{The parameter space of models we study in this work. At two ends $\beta=\pm 1$, there are two integrable models, and there is a special point $\beta=\frac{1}{3}$ at the middle, known as the AKLT model.}
    \label{fig:scheme}
\end{figure}

In this article, we investigate the model \eqref{ham} at a generic $\beta$ with the effective Bethe ansatz (EBA) approximation proposed in \cite{wenlong}. The key idea is that we still assume that the wavefunction takes the same form, i.e. the Bethe wavefunction, but the Bethe roots $u_i$ and $v_j$ are no longer solutions to the Bethe ansatz equation like \eqref{BAE}. Instead, we determine the effective Bethe roots $u_i$ and $v_j$, which shift from their values at the integrable point due to the perturbation deformation, by optimizing a proper loss function. Of course, at the integrable points,  the method exactly recovers the Bethe ansatz solution. This variational approach falls within the framework of the variational quantum eigensolver (VQE, refer to the review \cite{VQE-review}), as investigated in \cite{Nepomechie:2021jak,Raveh:2024cfh}.

Let us summarize the optimization algorithm below: 
\begin{enumerate}
    \item Prepare an off-shell Bethe state $\ket{\varphi_0}=\ket{\overline{\psi_{M_0}(\vec{u})}}$ or $\ket{\varphi_0}=\ket{\overline{\Psi_{N_0,K_0}(\vec{u},\vec{v})}}$ (depending on which integrable point to start with). 

    \item Find the value of Bethe roots $\vec{u}^{(0)}$ and $\vec{v}^{(0)}$ for the ground state by optimizing the loss function $L_0=\bra{\varphi_0}{\cal H}\ket{\varphi_0}$. We then obtain the ground state wavefunction, $\ket{\varphi_0}=\ket{\overline{\psi_{M_0}(\vec{u}^{(0)})}}$ or $\ket{\varphi_0}=\ket{\overline{\Psi_{N_0,K_0}(\vec{u}^{(0)},\vec{v}^{(0)})}}$. 

    \item Find the wavefunction of the 1st excited state by optimizing $L_1=\bra{\varphi_1}{\cal H}\ket{\varphi_1}+\lambda\left|\langle \varphi_1\ket{\varphi_0}\right|^2$. The second term forces the new state we want to find is orthogonal to the ground state $\ket{\varphi_0}$, and $\lambda$ is chosen large enough (we typically set $\lambda=100$). The wavefunction of the 1st excited state is then given by $\ket{\varphi_1}$.

    \item Repeat step 3 to find the $i$-th excited state by optimizing the Loss function $L_i=\bra{\varphi_i}{\cal H}\ket{\varphi_i}+\sum_{j=1}^{i-1}\lambda\left|\langle \varphi_{1}\ket{\varphi_{j}}\right|^2$, until we identified all the states we are interested in. 
\end{enumerate}
We note that in addition to the Bethe-root parameters $\vec{u}$ and $\vec{v}$, the magnon numbers $M_i$ (or $(N_i,K_i)$) for each state also highly depend on the model. In principle, we shall scan all the possibilities of magnon number to find each state in order. However, since the state of $M$ magnons is orthogonal to any state of $M'$ magnons for $M\neq M'$, one can also focus on each magnon sector and search for the state we are interested in. 

In this article, we focus on the anti-ferromagnetic ground state and the 1st excited state of the Hamiltonian \eqref{ham}. As depicted in Figure \ref{fig:scheme}, we start from both integrable ends ($\beta=\pm 1$) to check whether the effective Bethe ansatz method allows us to explore the non-integrable bulk of this family of models. A particularly interesting point appears at $\beta=\frac{1}{3}$, where the ground state can be constructed rigorously from a projection operator \cite{Affleck:1987vf,Affleck:1987cy}, and it is often called the Affleck-Kennedy-Lieb-Tasaki (AKLT) model. 

Around $\beta=-1$, we apply the (effective) rank-1 Bethe ansatz method, and the ground state and the 1st excited state respectively locate at $M=L$ and $M=L-1$. From the other end (around $\beta=1)$, we shall use the nested Bethe ansatz to approximate the wavefunction. In this region, as we will discuss later, the ground state is given by a superposition of states with different magnon numbers (depending on $L$), but it still gives a good approximation to take $(M_1,M_2)=\left(\frac{L}{2}+1,\lfloor\frac{M_1}{2}\rfloor\right)$ for the ground state. Similarly, we will focus on the sector $(M_1,M_2)=\left(\frac{L}{2},\lfloor\frac{M_1+1}{2}\rfloor\right)$ to study the 1st excited state. 

\begin{figure*}[htpb]
  \centering
  \begin{subfigure}[b]{0.7\textwidth}
    \includegraphics[width=\textwidth]{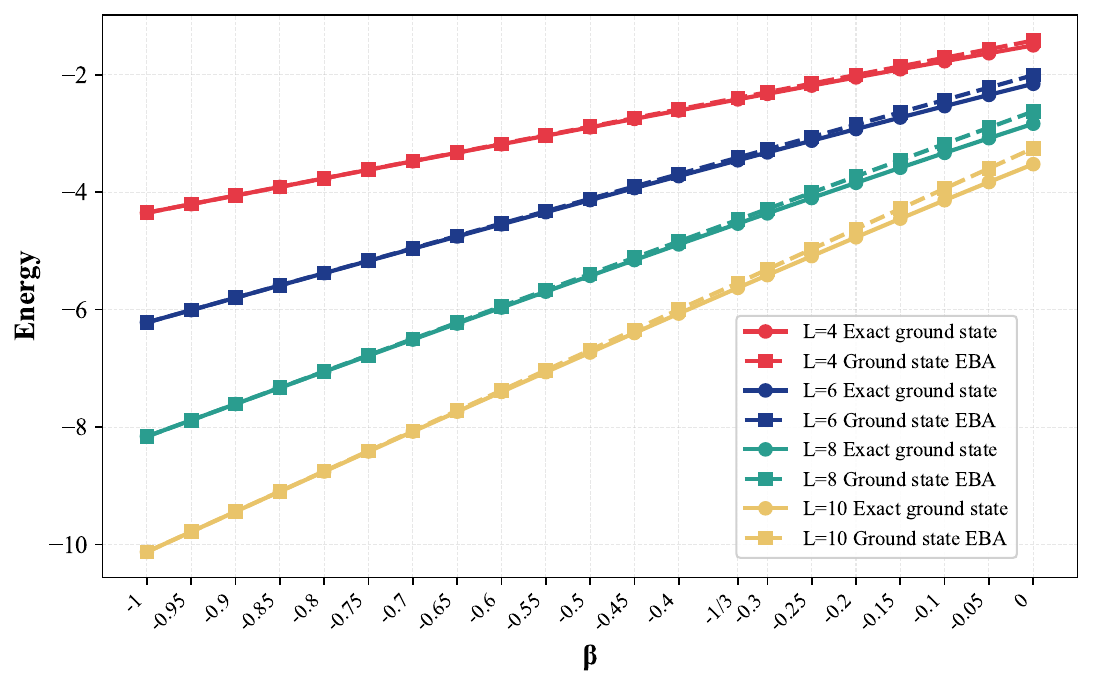}
    \caption{The comparison between the results found by ED (exact) and EBA (optimized) for the ground state.}
    \label{fig:gsm1}
  \end{subfigure}
  \vspace{0.5cm}
  \begin{subfigure}[b]{0.7\textwidth}
    \includegraphics[width=\textwidth]{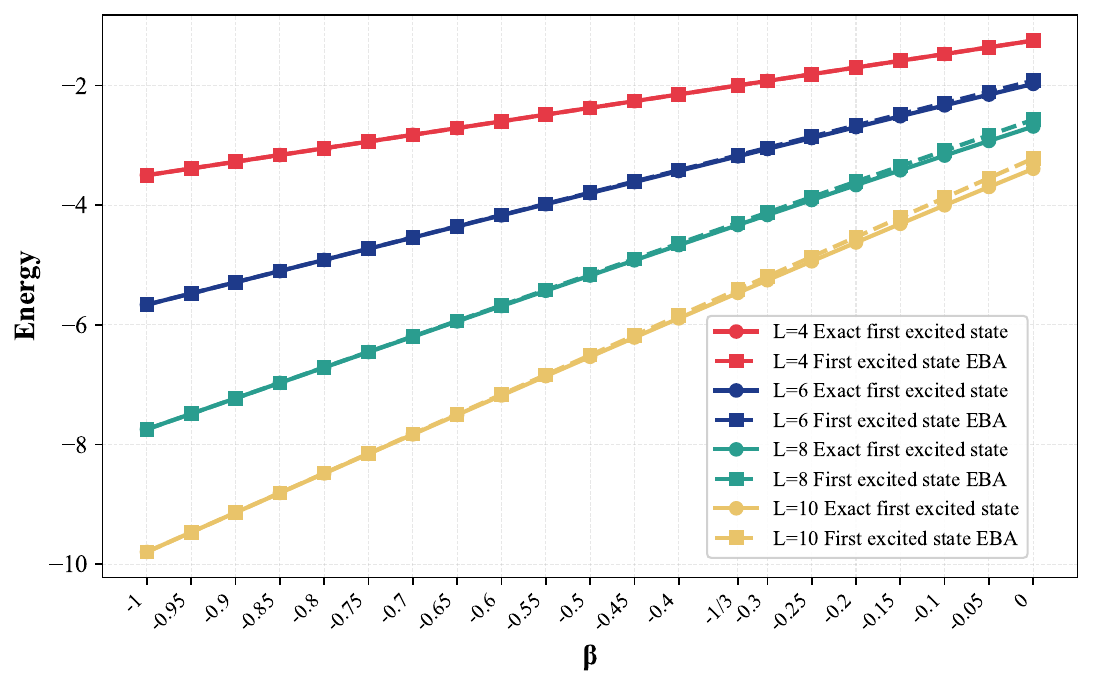}
    \caption{The comparison for the 1st excited state.}
    \label{fig:esm1}
  \end{subfigure}
  \vspace{0.5cm}
    \begin{subfigure}[b]{0.7\textwidth}
    \includegraphics[width=\textwidth]{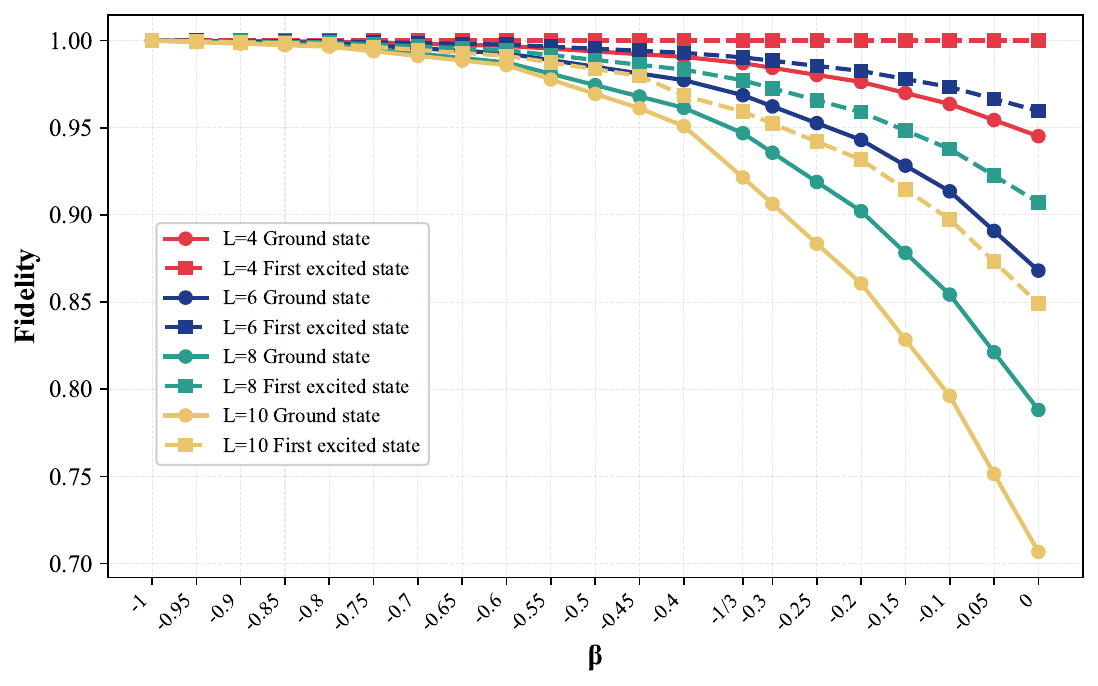}
    \caption{The fidelity found by EBA. }
    \label{fig:fidm1}
  \end{subfigure}
  \caption{}
\end{figure*}

\begin{figure}
    \centering
    \includegraphics[width=0.7\linewidth]{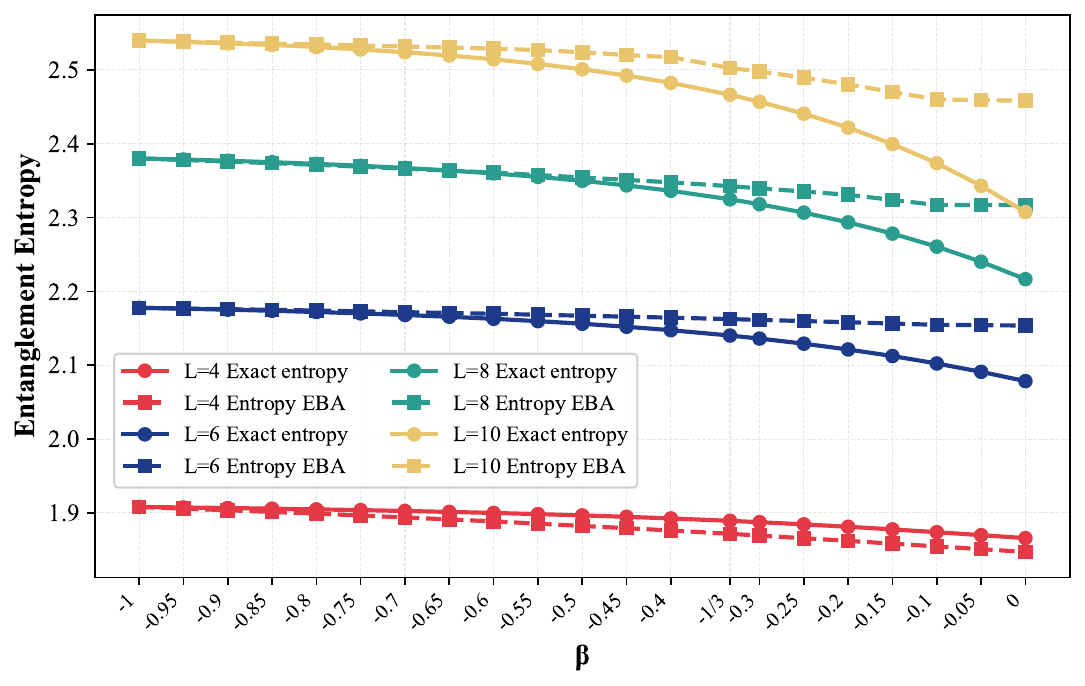}
    \caption{Comparison between the entanglement entropy found by ED and EBA.}
    \label{fig:EE-m1}
\end{figure}

\section{Results}\label{s:result}

In this section, we summarize the results obtained from the effective Bethe ansatz (EBA) applied to the Hamiltonian \eqref{ham}. We consider periodic chains of length $L=4,6,8,10$ and focus on the antiferromagnetic ground state and the first excited state.

\begin{figure}
    \centering
    \includegraphics[width=0.75\linewidth]{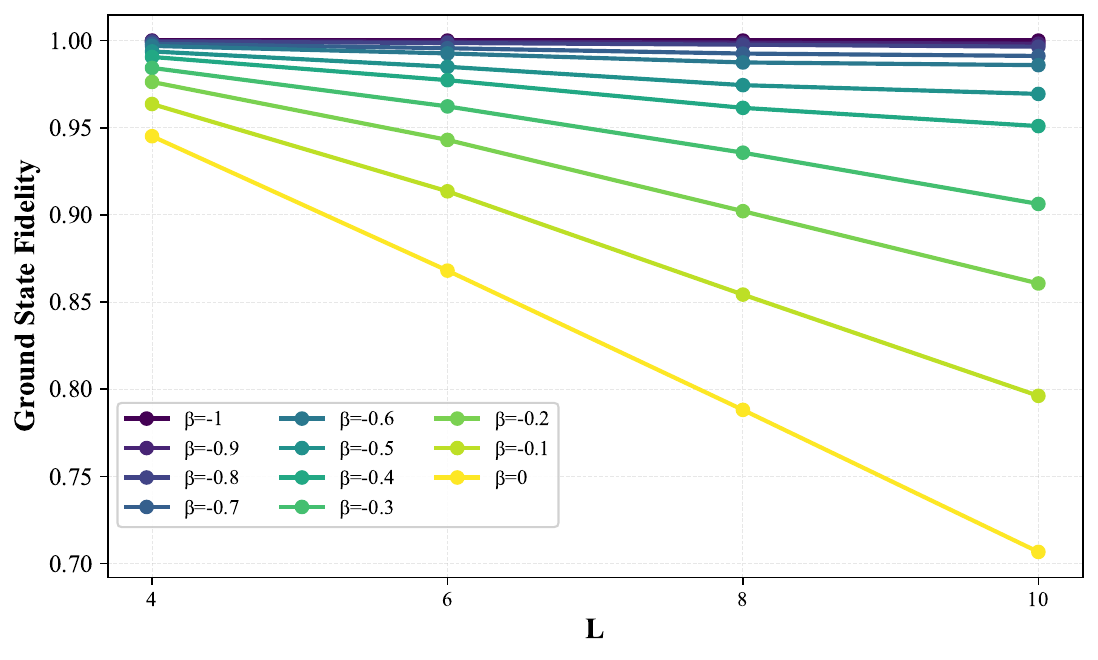}
    \caption{Fidelity vs Spin Chain Length. The fidelity appears to decay linearly.}
    \label{fig:fid-1}
\end{figure}

\begin{figure}
    \centering
    \includegraphics[width=0.75\linewidth]{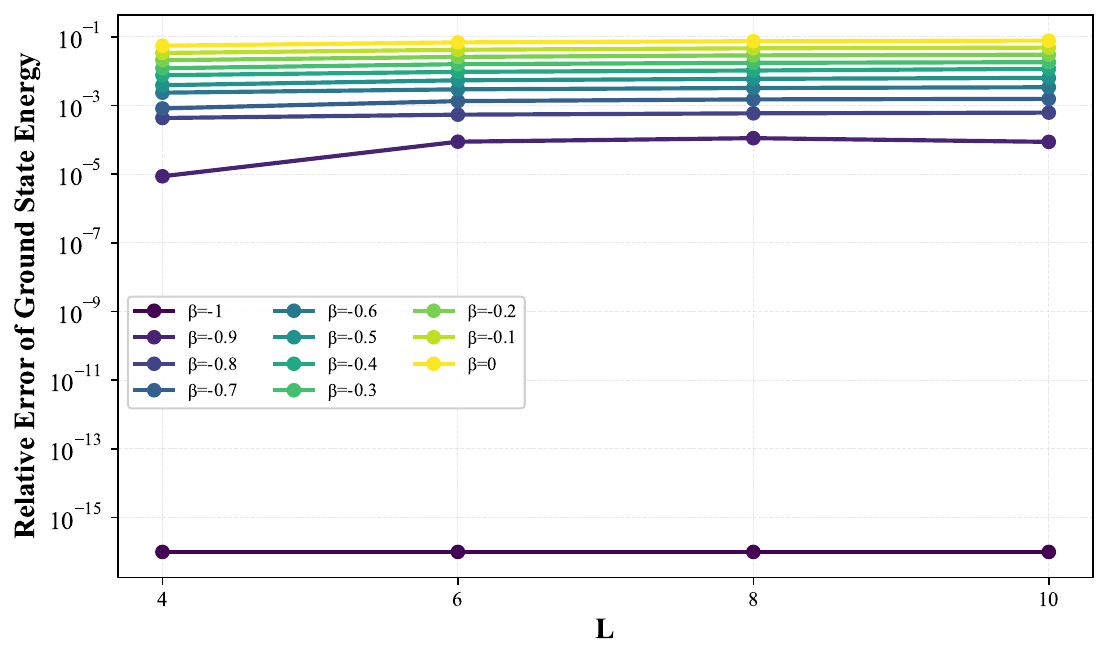}
    \caption{Energy error for ground state found by EBA.}
    \label{fig:error}
\end{figure}

\paragraph{From Takhtajan–Babujian point ($\beta=-1$):} The EBA method performs reliably until approximately $\beta\sim 0$. Figure \ref{fig:gsm1} and \ref{fig:esm1} respectively show comparisons between the energy levels obtained from exact diagonalization (ED) and those from the effective Bethe ansatz for the ground state and the first excited state. Fidelity decreases from $1$ to around $0.8$ as the perturbation from the integrable point grows large and $\beta$ approaches $0$ (see Fig. \ref{fig:fidm1}). Here, fidelity is defined as 
\begin{equation}
    \mathrm{Fidelity} = \frac{|\langle \psi_{\mathrm{ED}} | \psi_{\mathrm{EBA}} \rangle|^2}{\langle \psi_{\mathrm{ED}} | \psi_{\mathrm{ED}} \rangle \langle \psi_{\mathrm{EBA}} | \psi_{\mathrm{EBA}} \rangle}. 
\end{equation}
Additionally, fidelity exhibits a nearly linear decline with increasing system size $L$ (Fig. \ref{fig:fid-1}), but notably the energy error, ${\rm Error}=\left|\frac{E_{{\rm ED}}-E_{\rm EBA}}{E_{\rm ED}}\right|$, almost does not change with $L$ (see Figure \ref{fig:error}). Based on our observations, the EBA yields more accurate results in sectors with fewer magnons, which explains why the excited state ($M=L-1$) is better described than the ground state ($M=L$).

We have also computed the entanglement entropy for both the ground state and the first excited state, comparing the EBA outcomes with exact ED results (Fig. \ref{fig:EE-m1}). The entanglement entropy predicted by EBA increasingly deviates from the exact value as $\beta$ departs from $-1$, and notably, its variation with $\beta$ is slower than that of the exact result.

The above results align with our anticipation, and the observations reported in \cite{wenlong}, since the effective Bethe ansatz becomes less accurate as the system deviates further from the integrable point.

\begin{figure}
    \centering
    \includegraphics[width=1\linewidth]{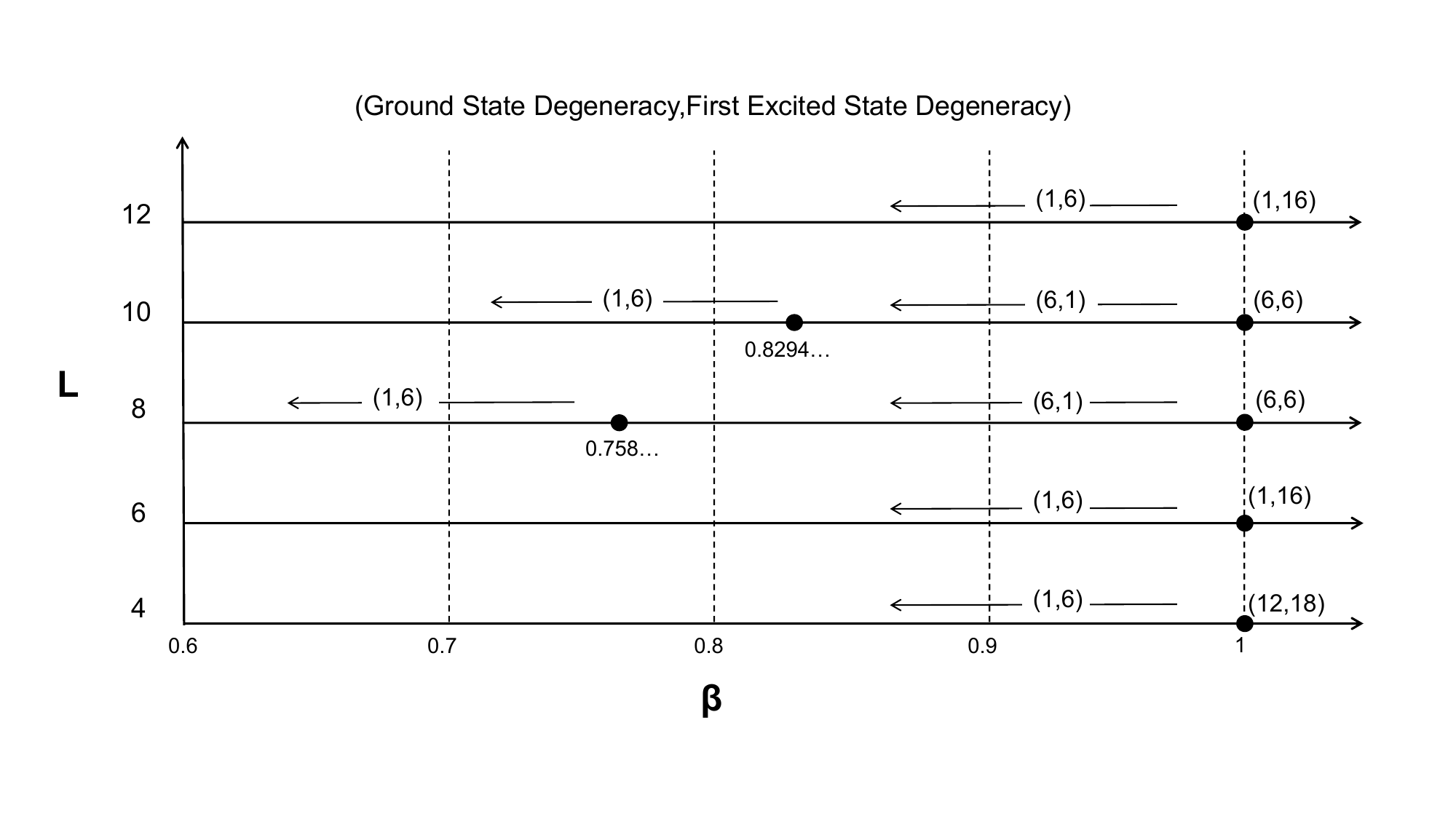}
    \caption{Degeneracy of the ground state and the 1st excited state for spin chains of different $L$. }
    \label{f:deg}
\end{figure}

\paragraph{From Lai–Sutherland point ($\beta=1$):} This region exhibits much richer physical behavior. In contrast to the unique ground state in the parameter range $\beta\in[0, 1]$, the degeneracy of the ground state and the first excited state highly depends on $L$ near $\beta=1$ (see Figure \ref{f:deg}). In the spin chain with length $L=4$, the antiferromagnetic ground state is unique for all the values of $\beta$ except for the integrable point $\beta=1$ with degeneracy $12$. For $L=6$, the ground state is always unique (including $\beta=1$), while the degeneracy of the first excited state reduces from $16$ to $6$ after moving away from the integrable point $\beta=1$. For $L=8$, the ground state has $6$-fold degeneracy when $0.75\lesssim \beta\leq 1$, and there is a unique 1st excited state in the same region. A level crossing between the ground state and the 1st excited state occurs at $\beta\sim 0.75$, and the ground state becomes unique when $\beta\lesssim 0.75$.

\begin{figure}
    \centering
    \includegraphics[width=1\linewidth]{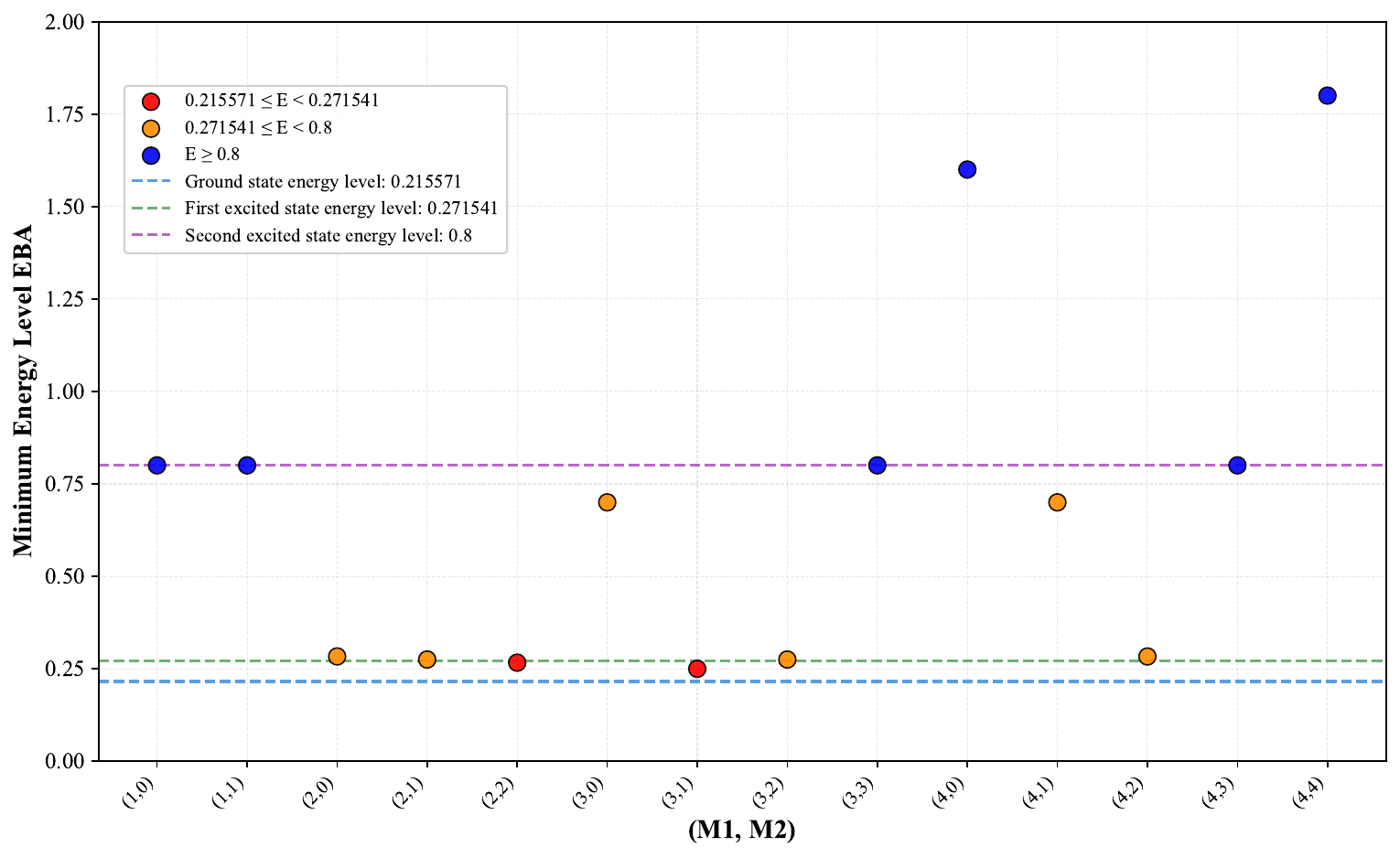}
    \caption{Lowest energy levels found by EBA in each magnon sector in the $L=4$ spin chain at $\beta=0.8$.}
    \label{pic1:5}
\end{figure}

\begin{figure}
    \centering
    \includegraphics[width=1\linewidth]{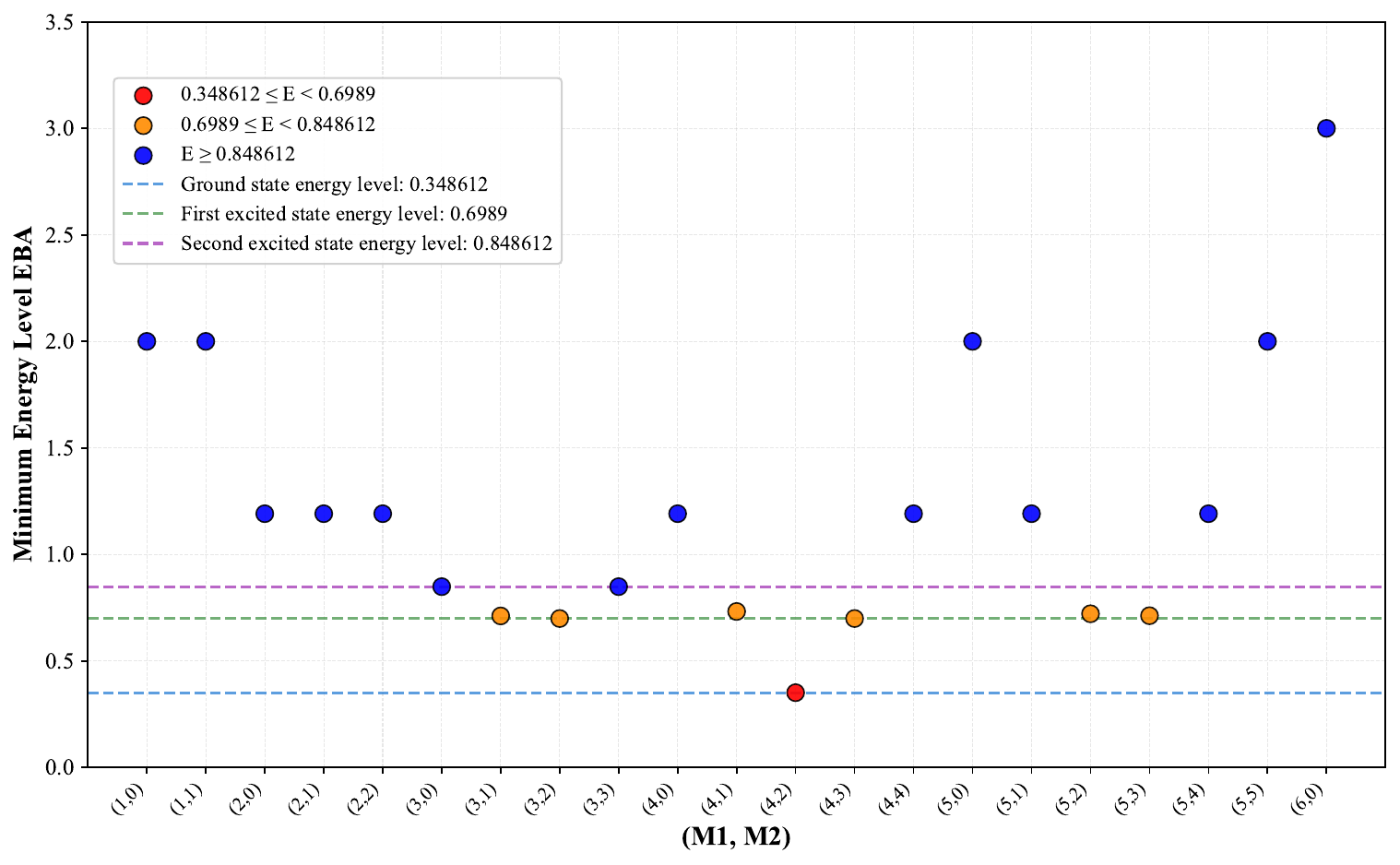}
    \caption{Lowest energy levels found by EBA in each magnon sector in the $L=6$ spin chain at $\beta=0.8$.}
    \label{pic1:6}
\end{figure}

\begin{figure}
    \centering
    \includegraphics[width=1\linewidth]{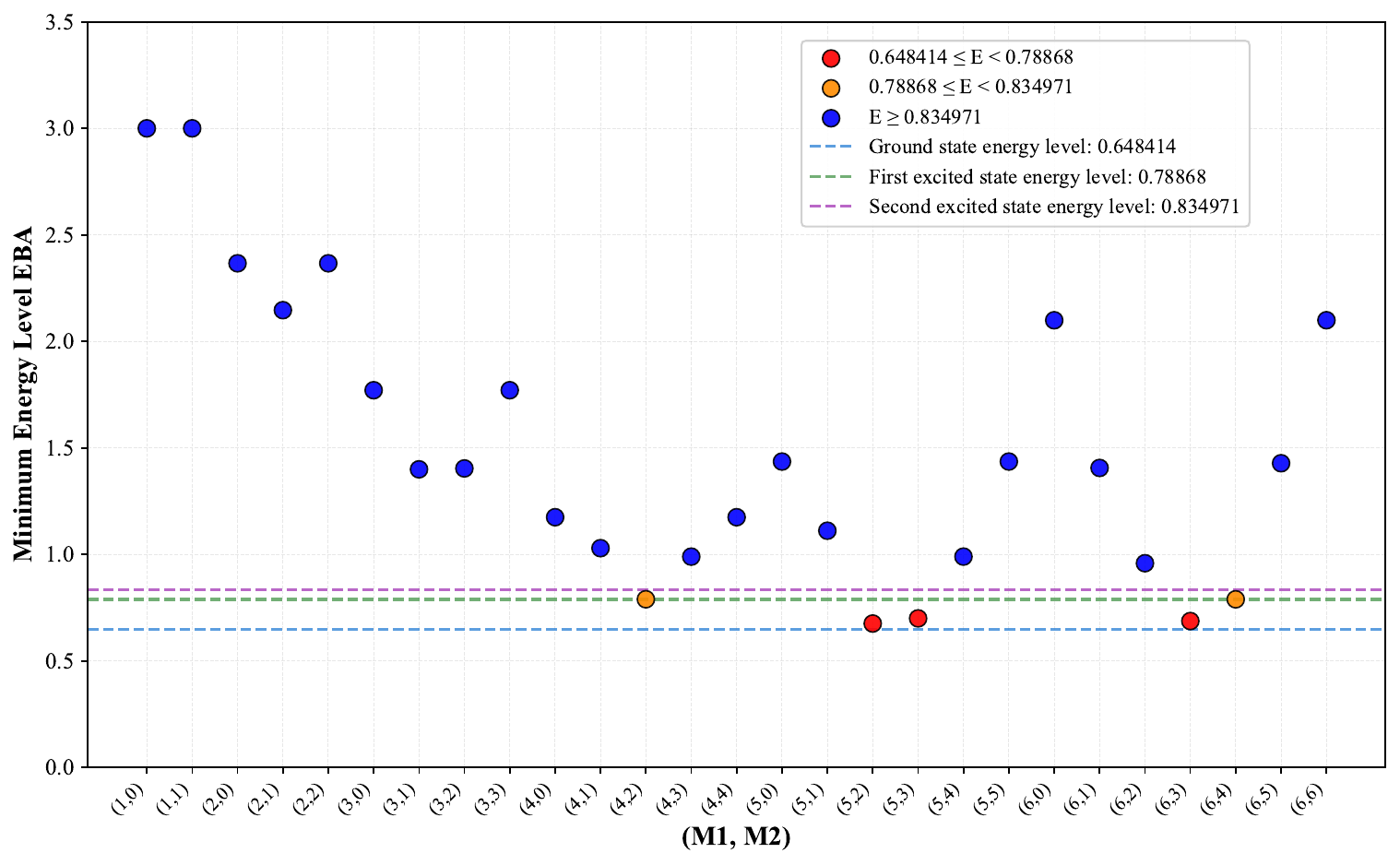}
    \caption{Lowest energy levels found by EBA in each magnon sector in the $L=8$ spin chain at $\beta=0.8$.}
    \label{pic1:7}
\end{figure}

\begin{figure}
    \centering
    \includegraphics[width=0.75\linewidth]{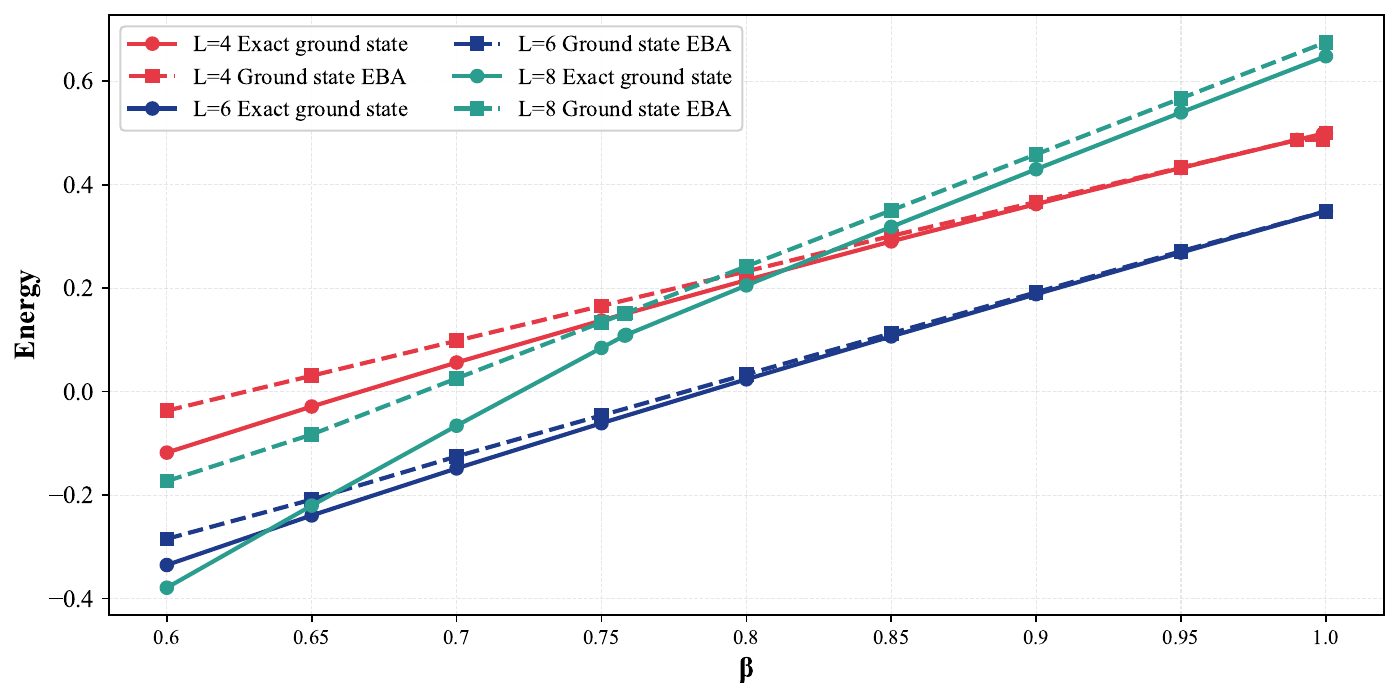}
    \caption{Ground-state energy compared between the results from ED (exact) and EBA (optimized).}
    \label{pic1:8}
\end{figure}

\begin{figure}
    \centering
    \includegraphics[width=0.75\linewidth]{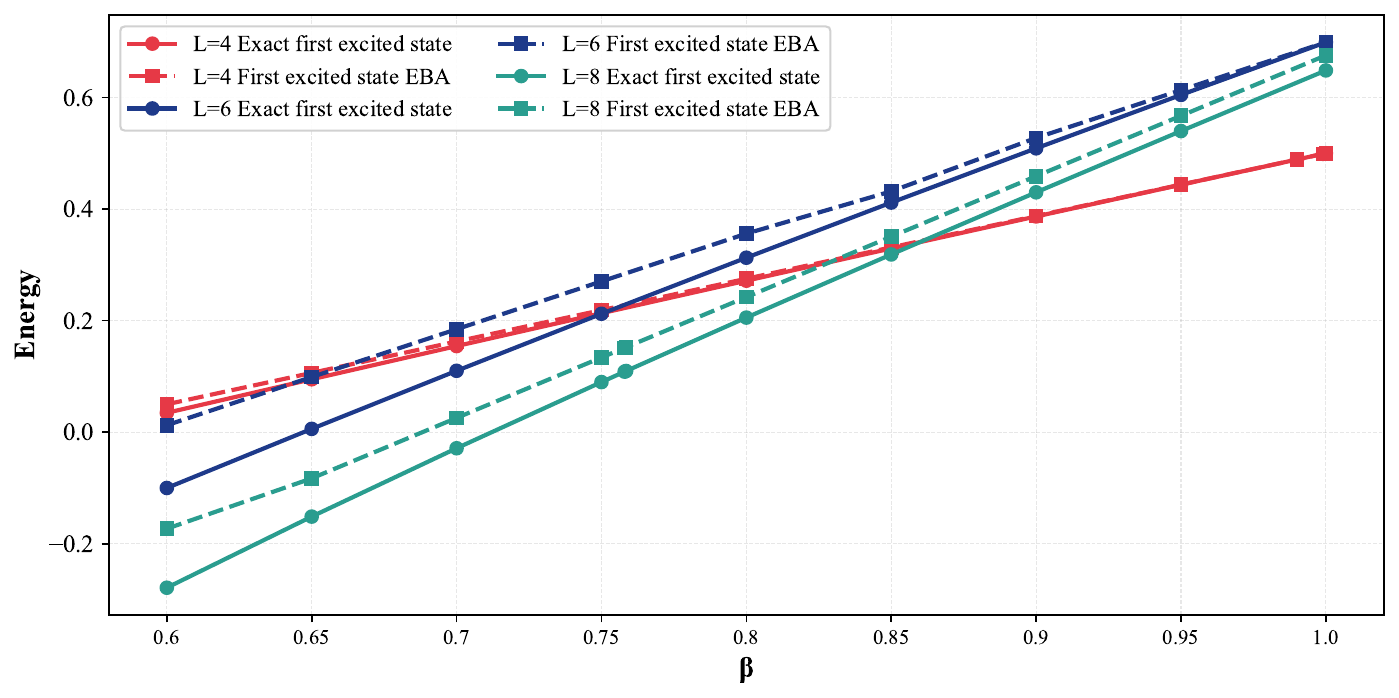}
    \caption{1st excited-state energy compared between the results from ED (exact) and EBA (optimized).}
    \label{pic1:9}
\end{figure}

\begin{figure}
    \centering
    \includegraphics[width=0.75\linewidth]{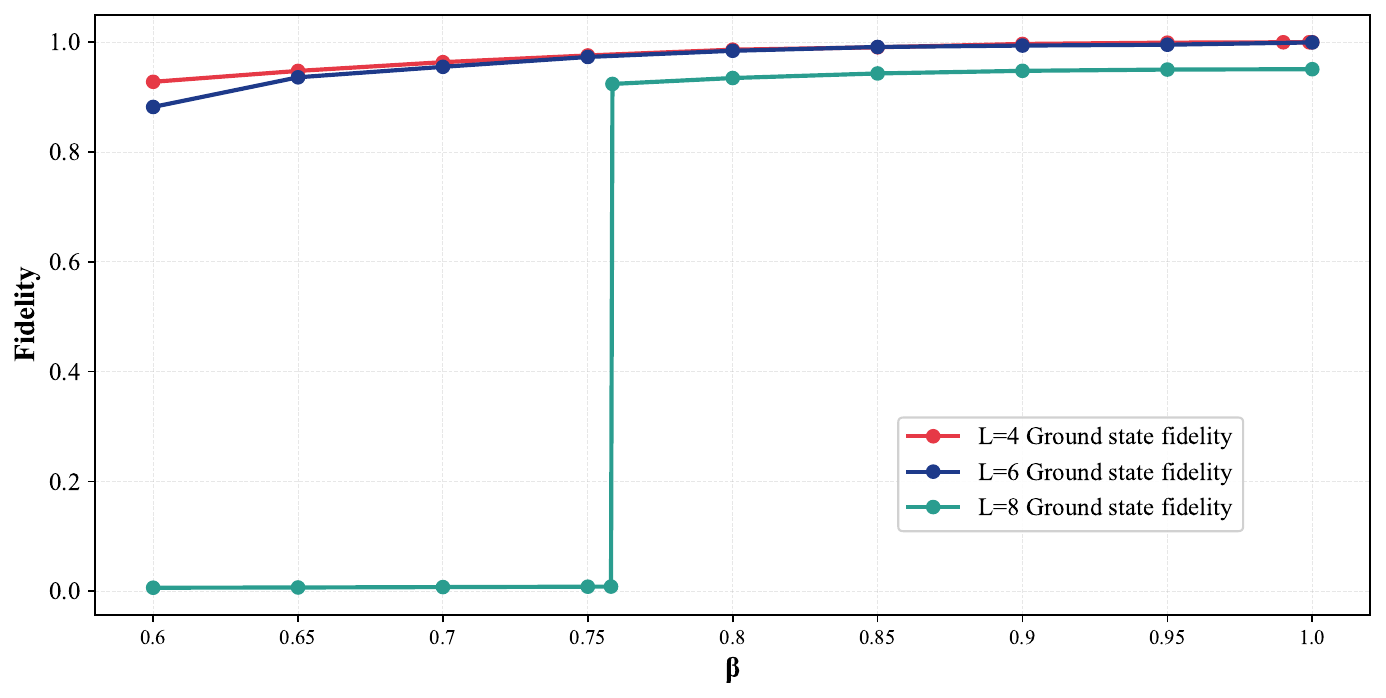}
    \caption{The fidelity of ground state found by EBA in spin chains with different lengths. $L=8$ case shows a sharp drop associated with the level crossing at $\beta\approx 0.7585$ (see also Figure \ref{f:deg}).}
    \label{fig:fid-gs-1}
\end{figure}

\begin{figure}
    \centering
    \includegraphics[width=0.75\linewidth]{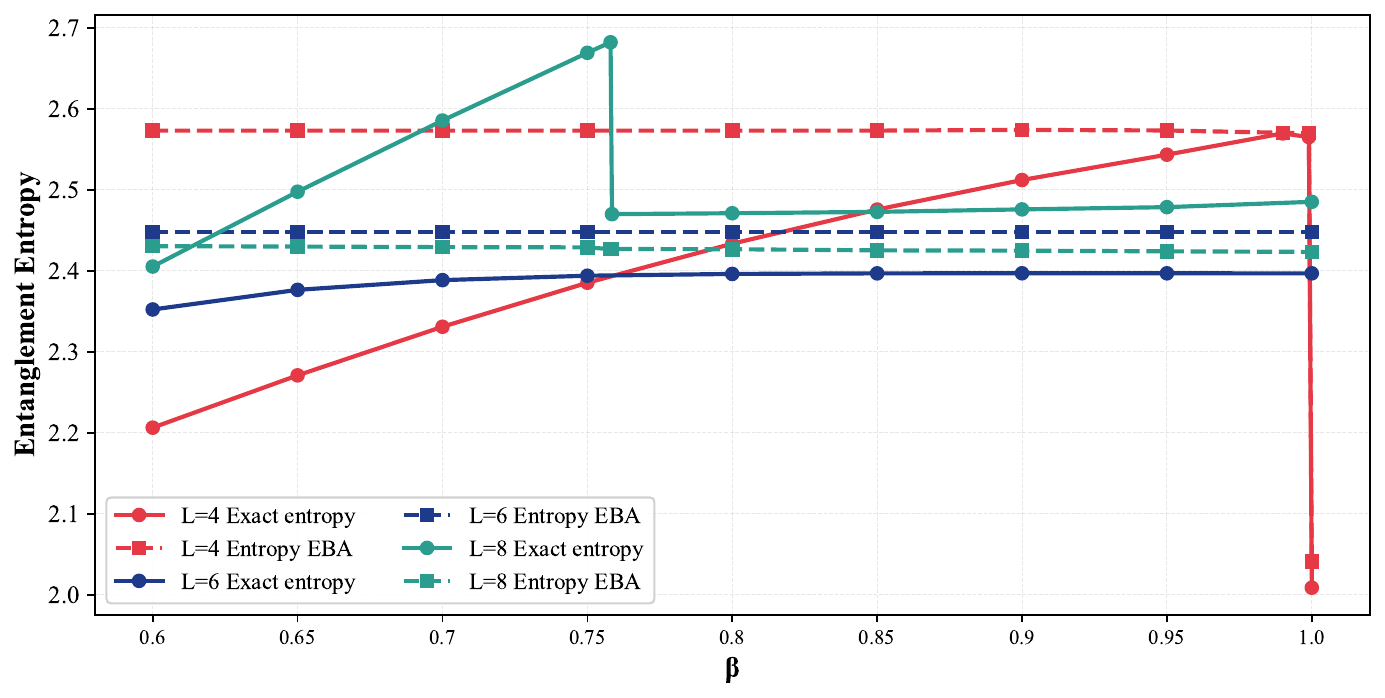}
    \caption{Comparison between the entanglement entropy found by ED (exact) and EBA. A sharp jump of the entanglement entropy in $L=8$ is again associated with the level crossing. }
    \label{fig:ee-1}
\end{figure}

We applied the EBA method to each magnon sector in the spin chains with $L=4,6,8$. In Figure \ref{pic1:5}, \ref{pic1:6} and \ref{pic1:7}, we respectively depicted the lowest-energy level in each $(M_1, M_2)$ sector obtained from EBA. As mentioned before, the lowest-energy level is found in the sector $(M_1,M_2)=\left(\frac{L}{2}+1,\lfloor\frac{M_1}{2}\rfloor\right)$. The 1st excited state, in general, is degenerate, so we simply pick up a pair $(M_1,M_2)=\left(\frac{L}{2},\lfloor\frac{M_1+1}{2}\rfloor\right)$ to access the effectiveness of the EBA method. In Figure \ref{pic1:8} and \ref{pic1:9}, we show the comparison between the energy level computed from EBA and ED, respectively for the ground state and the 1st excited state. The fidelity of the ground states found by EBA is presented in Figure \ref{fig:fid-gs-1}, and in Figure~\ref{fig:ee-1}, we compare the optimized results of the bipartite entanglement entropy of the ground state for the subsystem with length $l=L/2$ from the EBA method and those computed using ED.

In general, the accuracy of the EBA gets worse again when we move away from the integrable point $\beta=1$, but there are several interesting phenomena to be noticed. 

In the spin chain of length $L=4$, there are two optimized states with magnon numbers $(2,2)$ and $(3,1)$ lying between the energy level of the ground state and the 1st excited state. Each of them only has fidelity with the true ground state less than $0.7$ for $\beta\neq 1$, suggesting that a superposition needs to be considered. Indeed, by optimizing the superposition state $\ket{\psi}=\alpha\ket{\overline{\Psi_{2,2}}}+\gamma\ket{\overline{\Psi_{3,1}}}$ with $|\alpha|^2+|\gamma|^2=1$, the fidelity is improved to $1$ near the integrable point $\beta=1$ (see the plot in Figure \ref{fig:fid-gs-1}). 

This is, in fact, a realization of the Stark effect, i.e. linear combinations of degenerate states become the eigenstates under perturbations, in the effective Bethe ansatz approach. A similar phenomenon has already been observed in \cite{wenlong}. For the excited states, it is also desired to use the superposition ansatz, however, since it involves twice as many parameters to be optimized, it is much more expensive than the original EBA. Unless the optimized result is bad enough, we will not adopt the superposition ansatz for other states in this work. 

In the spin chain with length $L=8$, the ground-state fidelity drastically drops to zero at $\beta\sim 0.75$ (see Figure \ref{fig:fid-gs-1}). This abrupt drop in fidelity coincides precisely with a ground-state level crossing at the same parameter point. As depicted in Figure \ref{f:deg}, the ground-state degeneracy changes from 1 to 6 at $\beta\approx 0.7585$, indicating an exchange of quantum numbers between the ground state and the first excited state. Consequently, the bipartite entanglement entropy also exhibits a sharp jump at this crossing (see Figure \ref{fig:ee-1}), reflecting the sudden structural change in the wavefunction. This is not an isolated example in $L=8$, a similar level crossing with associated drops in fidelity and jumps in entanglement entropy is observed for $L=10$ at $\beta\approx 0.8294$. It is crucial to note, however, that these features are finite-size effects marking the crossing of two low-lying levels, and it is expected to converge toward $\beta=1$ as the system size increases. Therefore, they do not signify a quantum phase transition inside the bulk of $\beta<1$. The true quantum critical point, separating the gapped Haldane phase from the gapless critical phase, remains at the integrable Lai-Sutherland point, $\beta=1$ \cite{Itoi1997}. In this way, we see that the EBA approach provides a good probe to the level-crossing and potential phase transitions behind.

\begin{figure}
    \centering
    \includegraphics[width=0.75\linewidth]{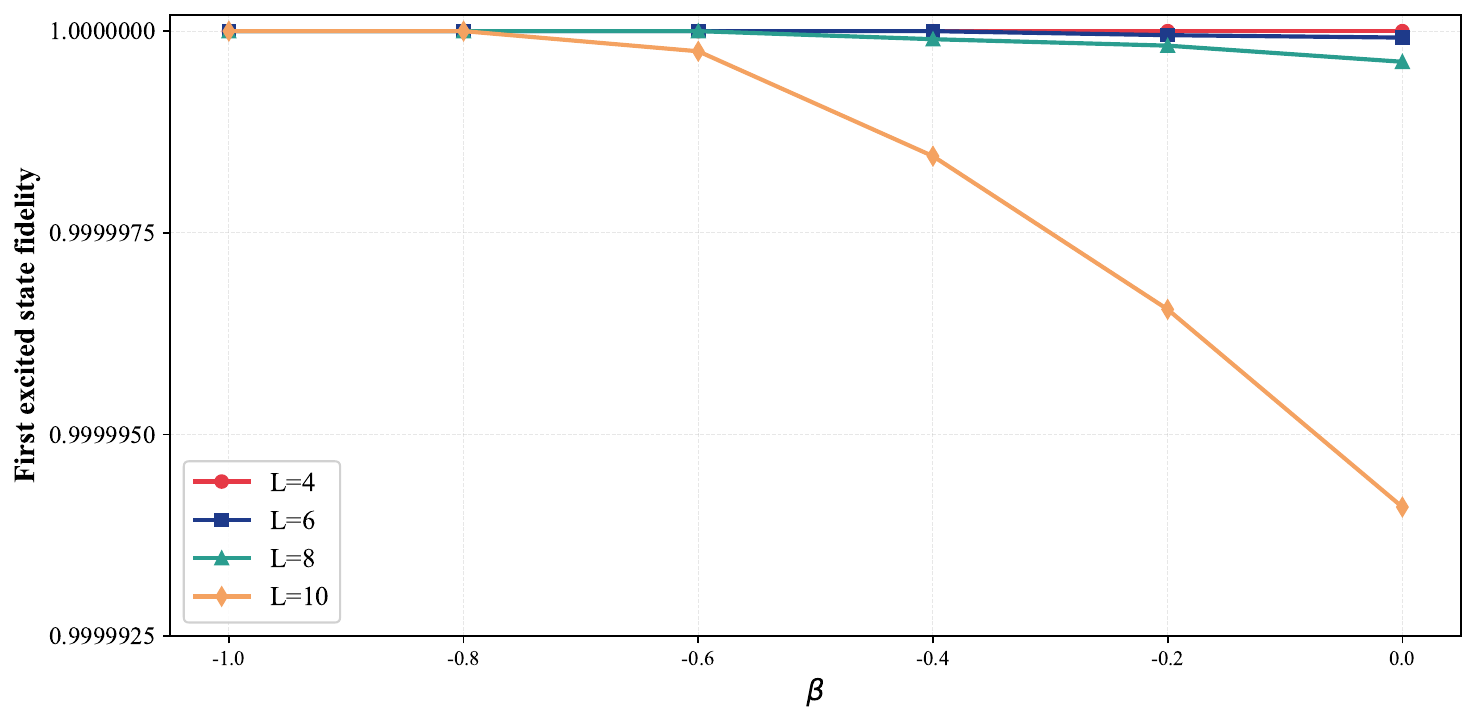}
    \caption{The fidelity of ground state of ferromagnetic system ($M=1$) found by EBA in spin chains with different lengths. All data are close to 1.}
    \label{fig:M=1}
\end{figure}

A similar jump of entanglement entropy is observed at $\beta=1$ in $L=4$ chain (see Figure \ref{fig:ee-1}). This is also due to the reduction of degeneracy in the ground state after turning on the perturbation, and the EBA approach manages to capture this feature. However, in the region $\beta<1$, it leads to a notable and counterintuitive observation: the optimized ground state achieves high fidelity, yet the corresponding entanglement entropy does not show the expected decreasing trend as one moves away from the integrable point. The effective Bethe ansatz fails to capture the detailed behavior of the entanglement entropy, which may be attributed to the constraints of its fixed wavefunction ansatz. This suggests that while EBA can approximate local observables and energy levels well, quantities that depend more sensitively on long-range entanglement may require an extended ansatz (e.g. superpositions of multiple Bethe sectors or additional variational parameters in the R-matrix).

\begin{figure}
    \centering
    \includegraphics[width=0.75\linewidth]{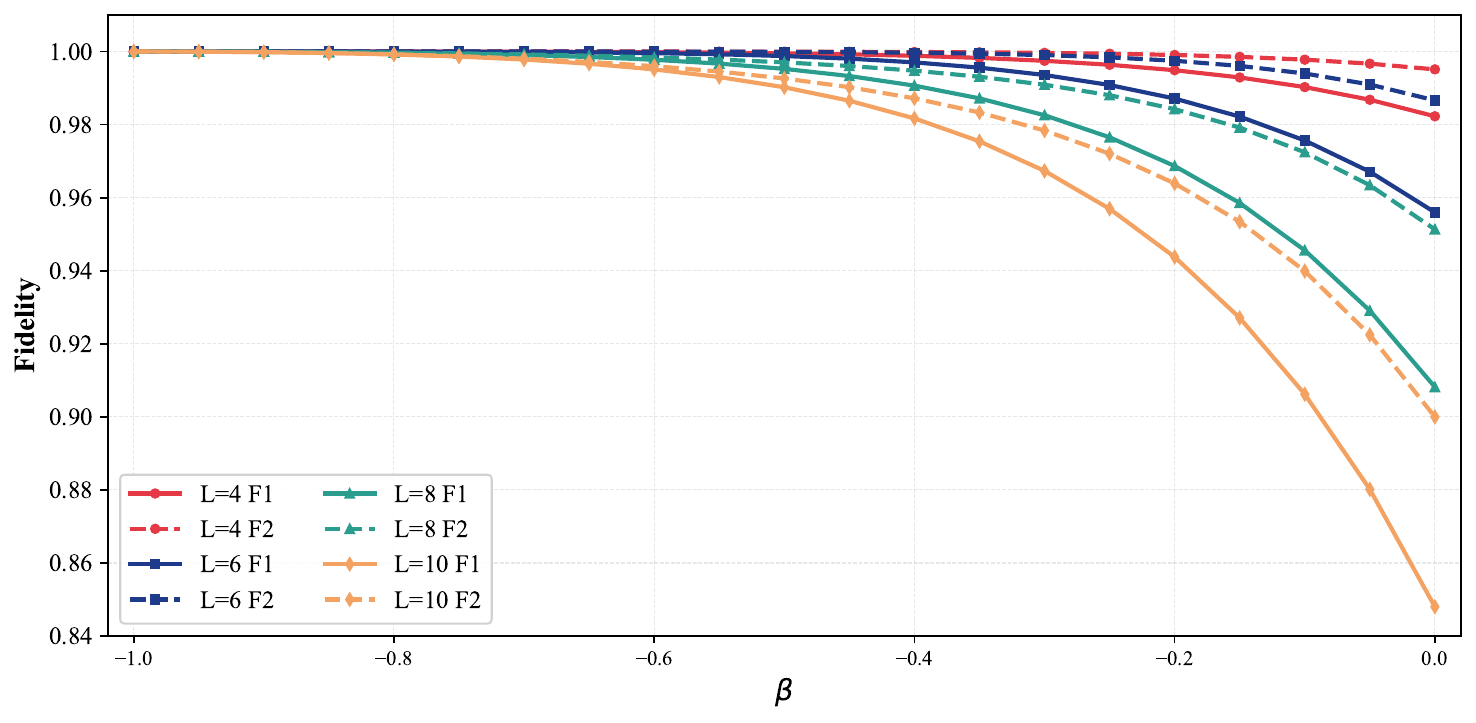}
    \caption{The ground state fidelity obtained via first order (F1) and second order (F2) perturbation theory starting from $\beta=-1$.}
    \label{1fidPT}
\end{figure}

\begin{figure}
    \centering
    \includegraphics[width=0.75\linewidth]{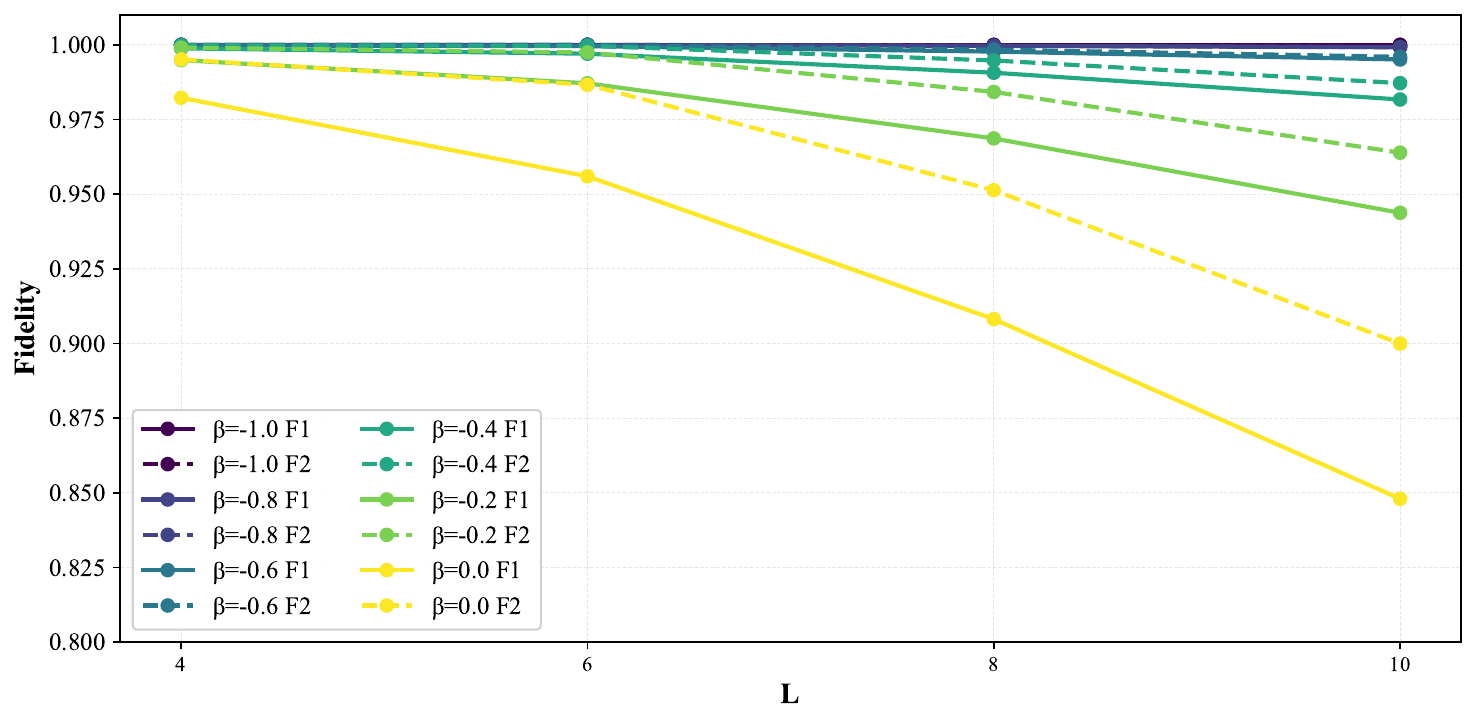}
    \caption{The dependence on $L$ of the ground state fidelity obtained via first order (F1) and second order (F2) perturbation theory starting from $\beta=-1$. The fidelity appears to decay slightly faster than linearly.}
    \label{1fidPTL}
\end{figure}

\begin{figure}
    \centering
    \includegraphics[width=0.75\linewidth]{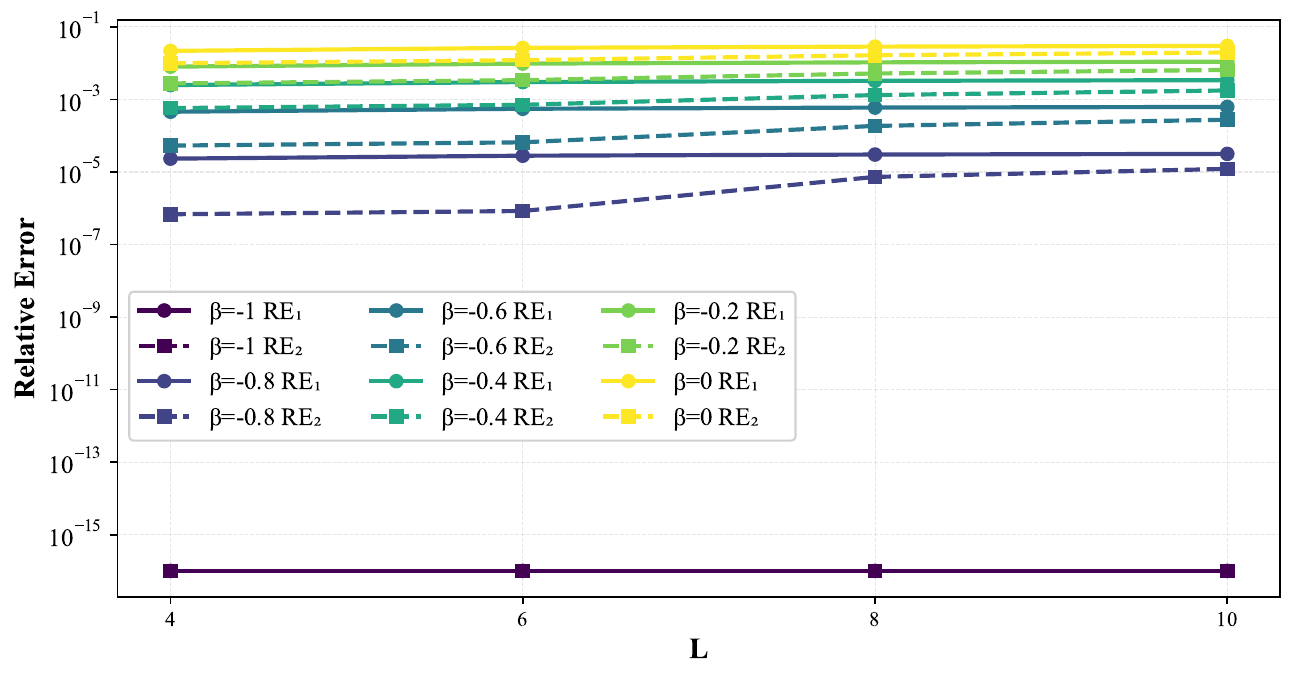}
    \caption{The relative error of the ground state energy obtained via first order (F1) and second order (F2) perturbation theory as a function of $L$ for $\beta$ ranging from $-1$ to $0$.}
    \label{1rePTL}
\end{figure}

\begin{figure}
    \centering
    \includegraphics[width=0.75\linewidth]{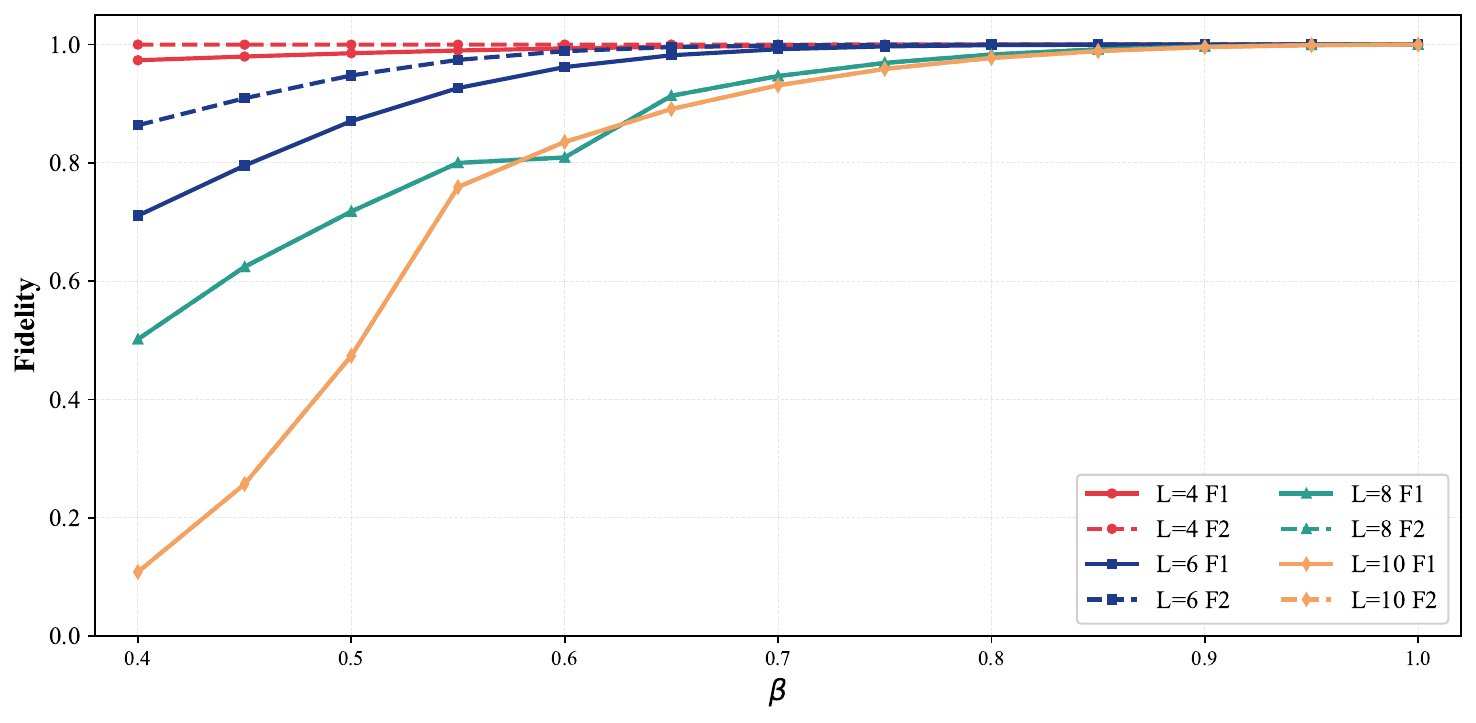}
    \caption{The ground state fidelity obtained via first order (F1) and second order (F2) perturbation theory starting from $\beta=1$.}
    \label{2fidPT}
\end{figure}

\begin{figure}
    \centering
    \includegraphics[width=0.75\linewidth]{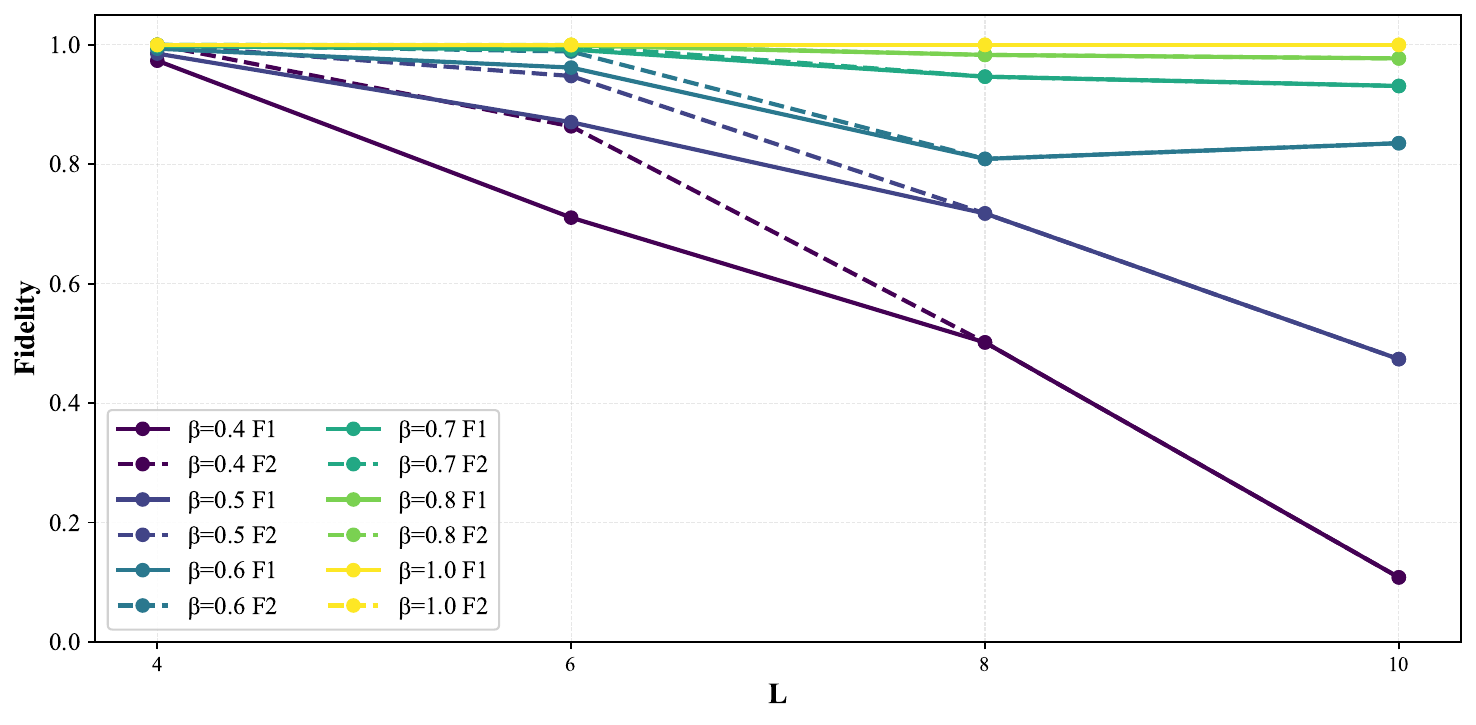}
    \caption{The dependence on $L$ of the ground state fidelity obtained via first order (F1) and second order (F2) perturbation theory starting from $\beta=1$.}
    \label{2fidPTL}
\end{figure}

\section{Comparison among EBA, ED, and Direct Application of Perturbation Theory}

In this section, we report the computational cost of the effective Bethe ansatz (EBA) algorithm and compare it with other traditional methods, such as exact diagonalization (ED) and perturbation theory. 

\paragraph{Computational scaling of ED and EBA} The principal computational bottleneck for both ED and the EBA approach is the exponential growth of the Hilbert space. However, the two methods differ qualitatively in how this exponential cost manifests. In ED, the Lanczos algorithm requires storing the Hamiltonian as a sparse matrix, and the memory complexity is estimated to ${\cal O}\left(L\binom{L}{M}\right)$, and in the computation of the anti-ferromagnetic ground state, it grows as ${\cal O}\left(L^{\frac{1}{2}}2^L\right)$. On a typical workstation with 16 GB of RAM, this limits ED of a spin-$1/2$ chain to $L\lesssim 24$ and a spin-$1$ chain to $L\lesssim 16$ at half filling. The EBA approach, by contrast, can be realized to construct the variational state $\ket{\psi_M(\vec{u})}$ and evaluate its energy expectation value entirely through matrix-free operations: each application of the $B$-operator and each local bond term acts on the state vector via sequential on-site manipulations, requiring no storage of the Hamiltonian. The memory footprint is therefore dominated by the state vectors alone: ${\cal O}\left(\binom{L}{M}\right)$ complex numbers. For $L=28$, the EBA algorithm requires only about $1/8$ of the memory needed by ED, and indeed, we managed to run the computation for spin-$1/2$ chain in the context of \cite{wenlong} at $L=26$ on a computer that cannot perform ED for the same system. 

The time complexity of ED is ${\cal O}\left(L^{\frac{3}{2}}2^L\right)$, and in the EBA algorithm, it depends on the iteration step $K$ to minimize the cost function, given by ${\cal O}\left(KL2^L\right)$. When we consider a system close enough to an integrable point, we can use the solution to the BAE as initial parameters, and it then only takes a few steps of $K$ to optimize. Notably, solving the BAE incurs only an additional ${\cal O}(M^3)$ time complexity, thus, the EBA approach is expected to outperform ED for sufficiently large $L$. 

In addition, once we have the analytic expression of the cost function\footnote{A brute-force way is to compute it for each $L$ and store it in our programme in advance. This will accelerate the computation by a lot when we want to scan the parameter space. }, $L=\bra{\psi_M(\vec{u})}H\ket{\psi_M(\vec{u})}$, the time complexity of the optimization process will be reduced to a polynomial level ${\cal O}(KM^4)$. At the integrable point, $L$ can be computed via the commutation relations in the algebraic Bethe ansatz approach along with the scalar product of off-shell Bethe states \cite{Korepin:1982gg,Slavnov:1989uvz,Liashyk:2025oao}, and it will be interesting to find a more clever way than the brute-force computation to give the analytic expression of the perturbation terms sandwiched by the off-shell Bethe state. 

\paragraph{Comparison with Perturbation Theory} Since our method shares the same spirit with the perturbation theory, it is natural to compare our results with it. In Figures \ref{1fidPT}, \ref{1fidPTL}, \ref{1rePTL}, \ref{2fidPT}, and \ref{2fidPTL}, we present the results computed from a program that performs 1st-order and 2nd-order perturbation analysis with the eigenstates at the integrable points ($\beta=\pm 1$) obtained by ED. Approaching from $\beta=-1$, a similar tendency of the fidelity dropping with $\beta$ (Figure \ref{1fidPT} vs \ref{fig:fidm1}) and $L$ (Figure \ref{1fidPTL} vs \ref{fig:fid-1}) is observed, and we note that the perturbation theory gives better results for the anti-ferromagnetic ground state than the EBA approach in the sense of fidelity. As for the ferromagnetic case with Hamiltonian, 
\begin{equation}
H=-\frac{1}{4}\left[\sum_{i=1}^L\vec{S}_i\cdot \vec{S}_{i+1}+\beta \left(\vec{S}_i\cdot \vec{S}_{i+1}\right)^2\right],\label{fham}    
\end{equation}
the EBA algorithm managed to identify one of the ground states in the $M=1$ sector (see Figure \ref{fig:fid-1}) and $M=2$ sector with fidelity extremely closed to $1$ in the region $0\geq \beta\geq -1$. The perturbation theory is quite involved in this case due to the large degeneracy, and gives fidelity around $\sim 0.9$. It is not surprising since the EBA is observed to work better with fewer parameters to be optimized. 

Turning on the perturbation from $\beta=1$ exhibits a rather different behavior from the EBA approach, e.g. no sharp drop is observed at the level-crossing point in the perturbation theory (see Figure \ref{2fidPT}). As already proposed in \cite{wenlong}, this confirms that the EBA method provides a new probe for level-crossing and quantum criticality in 1D near-integrable models. 

It is worth remarking that the proposed EBA algorithm should not be viewed merely as a classical numerical solver, as it remains susceptible to exponential scaling. We argue that its significance stems from two aspects: first, it maintains the underlying integrability structure, providing a valuable semi-analytic approach; second, it may pave the way for polynomial-time computation if successfully realized on near-future quantum hardware. More discussions will be presented in the conclusion section. 

\section{Conclusion and Outlook}\label{s:conclusion}

In this work, we have implemented and benchmarked the Effective Bethe Ansatz for the non-integrable spin-1 bilinear-biquadratic chain. By deforming the Bethe roots of the exact wavefunctions at the two integrable points ($\beta=\pm 1$), the EBA method provides a good approximation for the ground state and low-lying excitations in their vicinity. The accuracy, quantified by the fidelity with exact diagonalization results, naturally degrades as the system moves further into the non-integrable regime, consistent with expectation.

Our study reveals several key insights. First, the EBA faithfully captures important physical features, such as the finite-size level crossing between the ground and first excited states for $L=8$ and $L=10$ near $\beta=1$, which is associated with a sharp drop in fidelity and a jump in entanglement entropy. This confirms that the EBA can serve as a sensitive probe for spectral rearrangements and near-critical behavior. Second, we find that a simple single-ansatz state is sometimes insufficient, and a superposition of states from different magnon sectors is required to achieve high fidelity, effectively realizing a Stark effect within the variational framework. This highlights the flexibility of the EBA method and its connection to perturbation theory.

There are several promising directions to be further explored in the future. A natural extension is to introduce free parameters into the underlying R-matrix that produces the off-shell Bethe ansatz, \eqref{rank-1-ansatz} and \eqref{nested-ansatz}. This idea has already been explored in \cite{wenlong}, and introducing the 8-vertex R-matrix or inhomogeneous parameters indeed improves the fidelity close to $1$ for the two models studied there. It would be desirable to examine the same idea in spin chains with higher spin or higher-rank symmetry groups. Furthermore, it would be highly instructive to test the generality of the EBA by taking the integrable starting point beyond the spin models, based on e.g. the Lieb-Liniger \cite{Lieb:1963rt,Lieb:1963zz} or Gaudin-Yang models \cite{Yang:1967bm,1967.Gaudin.PLA.24} for continuum particle models, and the Fermi-Hubbard \cite{Lieb:1968zza} or unidirectional Bose-Hubbard models \cite{Zheng:2023jjd} for lattice systems. 

Our work so far has focused on spin chains with periodic boundary conditions, but exploring spin chains with open boundary conditions represents another crucial frontier, as it would connect the formalism directly to experimental realizations in finite systems and quantum simulators. The Bethe ansatz formalism was first extended to open chains with diagonal K-matrices by Sklyanin \cite{Sklyanin:1988yz}. It is straightforward to generalize the effective Bethe ansatz method to include diagonal K-matrices, and we plan to investigate it in future work. General open boundaries with non-diagonal K-matrices are more challenging, and it was first solved by the off-diagonal Bethe ansatz proposed in \cite{Cao:2013nza}. The wavefunction is given in a superposition form via the algebraic Bethe ansatz \cite{Belliard:2013aaa}, and it would be interesting to utilize the off-diagonal Bethe ansatz to investigate non-integrable models without U(1) symmetry. 

A systematic investigation of the thermodynamic limit is equally essential. Developing a robust scheme to extract and analyze the distribution of effective Bethe roots in this limit is particularly crucial, as it would directly reveal the underlying thermodynamic behavior of the system by mimicking the thermodynamic Bethe ansatz (TBA) \cite{Yang-Yang66,Takahashi1971}. A similar idea has been developed in \cite{doi:10.1143/JPSJ.63.3598} (see also \cite{2015PhRvB..92l5136V}) to write down a TBA equation for the BLBQ Hamiltonian based on the approximate Bethe ansatz method with two-body scattering matrix solved at each $\beta$, though it works well mainly in the low particle-density regime. The root distribution, which is observed to have some pattern in the EBA approach, encodes the macroscopic physics, and its analytical continuation from the integrable points could provide a powerful framework for understanding properties like excitation spectra and correlation functions in the non-integrable regime, thereby bridging the microscopic pseudo-particle description with field-theoretical predictions. We leave it to future work, as it is way beyond the scope of the current paper. 

Finally, exploring the synergy between the Effective Bethe Ansatz and quantum computing presents a forward-looking direction. Compared to generic variational circuits, the EBA ansatz contains fewer variational parameters and encodes integrability-based physical intuition. Encoding the Bethe ansatz on a quantum processor could enhance VQE approaches for lattice models, potentially offering advantages in convergence speed and resilience to noise compared to more generic circuits. This prospect is supported by recent quantum circuit implementations of the coordinate Bethe ansatz \cite{Dyke:2021vkq,VanDyke:2021nuz,Li:2022czv,Raveh:2024llj} and the algebraic Bethe ansatz \cite{Sopena:2022ntq,Ruiz:2023rew,Ruiz:2025qmt}, to which the variationally optimized effective Bethe roots can be directly transcribed. This may open the path to studying quantum many-body dynamics on quantum simulators. On a quantum device, the exponential overhead of classical EBA may be circumvented: mapping each spin-$1/2$ site to a qubit (with spin-$1$ site embeded into the subspace of two qubits), the Bethe state is estimated to be prepared by a  parametrized circuit of depth ${\cal O}(ML)$ \cite{Ruiz:2025qmt} using only $L$ qubits for spin-$1/2$ chains, and energy gradients are computed exactly via the parameter-shift rule in $2M$ circuit evaluations. This reduces the per-iteration cost from $O(KL 2^L)$ (classical) to $O(KML)$ measurements, an potential exponential speedup in the system size $L$. By preparing approximate eigenstates obtained via the EBA and simulating real-time evolution, one could probe spectral properties, dynamical correlation functions, and thermalization dynamics in non-integrable regimes, thereby bridging approximate analytical insights with the programmable capabilities of quantum simulation platforms.

\bibliography{QC}
\bibliographystyle{JHEP}

\end{document}